\documentclass{article}
\usepackage[]{fontenc}
\title{Gravity Role in Classical Electrodynamics of Charged Point Source }
\author{M.B.Golubev\footnote{e-mail: mikhailg@albatross.md08.vniief.ru}}
\date{}
\begin{document}
\maketitle
{\begin{center}
Russian Federal Nuclear Center - All-Russian S\&R Institute of
Experimental Physics

\end{center}}
\begin{abstract}
This paper deals with the problem of a point-like charged source under the influence of the external electromagnetic field in terms of perturbation theory for GR equations. It is obtained that GR, in contrast with the classical electrodynamics, in linear perturbation theory predicts an unlimited growth of the dipole perturbation. It is shown that the reason for this unlimited perturbation growth might be related to the presence of the unstable rotational perturbation mode. The analysis of the conditions under which this instability may disappear is performed. The momentum value at which the stability is reached is estimated. These estimations give the electron spin  by the order of magnitude (when charge value is equal to elementary one).
\end{abstract}
\section*{Introduction}
The gravitation force is commonly considered negligible in
elementary particle interaction since proton-electron gravitation
interaction force is forty orders of magnitude less than that of
electromagnetic interaction. On the other hand, Einstein's equation
system is closed, while Maxwell's equations requiring closing by sources
motion equations. In other words, within the framework of general
relativity it is possible to pose a problem of an electromagnetic
field source (a charged point source or a charged black hole) motion and emission
under the influence of an external electromagnetic field. So while
it is necessary to postulate Lorenz's force in electrodynamics, it
is possible to obtain it using general relativity.

This statement does not look like a paradox if one takes into
account the following fact. It is possible to postulate in
electrodynamics not Lorenz's force but the energy conservation law.
In this case it is possible to obtain Lorenz's force in
electrodynamics as well. Einstein's equations provide for energy
conservation automatically because of Bianchi's identity. So it is
not at all surprising that they should also contain field sources
equations of motion (Lorenz's force). However, there is no ground
to expect {\em a priori} that a field source equations of motion will be
the same in electrodynamics and in GR. More than that, physical
mechanisms leading to a source acceleration are basically different
under electrodynamics and GR. In electrodynamics the acceleration
is a result of electromagnetic field interaction with the charge
(by Lorenz's force). Since for a pointlike charge the field at the
point of charge has infinite intensity, Lorenz's force for a pointlike 
charge is ill - defined and requires a renormalization. As a
result, non-physical runaway solutions exist in
electrodynamics. In GR the acceleration emerges as a result of the
external field interaction with the source own field in some
vicinity of the source. This interaction contributes to the
electromagnetic field energy resulting in metrics curvature that is
interpreted as the acceleration by an external observer.

Since the right-hand side of Einstein's equations $G^{\mu\nu}=\kappa T^{\mu\nu}$
 contains the
gravitation constant as a multiplier to the interaction energy
while the left-hand side contains the metric tensor derivatives
(including acceleration), one may agry that
 the acceleration should then be proportional to the gravitation
constant. Let's show how a small quantity - gravitation constant -
falls out from the expression for the acceleration. Let $g^0_{\mu \nu}+h_{\mu \nu}$
 be a charged source metrics perturbed by the external field. 
The main term of the nonperturbed part  is $g^0_{00}=1-\kappa m 
c^2/r+\kappa Q^2/r^2$ where $m$ is the mass and $Q$ is the charge.
The main term of the perturbed part  is $h_{00}=2 \vec a \vec r /c^2$,
 where $\vec a$ is the acceleration. The
main term by the powers of $r$ in the expression for the linear by
the perturbation part of the Einstein tensor will be: $\partial 
g^0_{00} \partial h_{00}\sim \kappa m c^2/r^2\quad a/c^2$. The corresponding term of the right-hand side of the linearized Einstein's equations
constitutes the energy of the external field interaction with the
source own field multiplied by gravitation constant: $\kappa 
T_{00}\sim \kappa Q/r^2 E$. So the
expression: $\kappa m c^2/r^2\quad a/c^2 \sim 
\kappa Q/r^2 E$ results in $a \sim Q E/m$.

So it is clear it's sufficient to consider the first order
perturbation theory equations in order to study a charge motion
dynamics in GR. One of most advanced perturbation techniques for GR -
Einstein-Infield-Hofman's procedures was used by
 Anderson~\cite{Ander} To obtaine not only Lorenz's force
but radiation reaction force as well. However, Einstein-Infield-
Hofman's technique uses decomposition by the powers of a few
parameters one of them usually being the charge. This limits
the above technique to the distances much greater
than the classical radius ($Q^2/mc^2$), since at the classical radius the
contribution of the charge into the metrics becomes equal to that
of the mass. Below we will show that at this distances of the
order of classical radius GR gives the results that are
substantially different from those of classic electrodynamics.
Bicak~\cite{Bicak} in 1980 studied the problem of a charged black hole
motion in a constant asymptotically uniform electric field using
perturbation theory linear by the amplitude. Acceleration obtained
by Bicak was exactly the same as the Lorenz's force. This paper
and paper~\cite{Aut} extends Bicak's technique to a charged point source
and time-dependent spatially non-uniform perturbation.

\section*{
Problem Statement and Gage-Invariant Perturbation Theory
Equations}

Let's place coordinate center at the charge center and consider
axially symmetric perturbations of the first order by amplitude
magnitude. Axially symmetric perturbations for spherically
symmetric metrics  as shown by Regge and Wheeler~\cite{Wheel} permit
variables separation, i.e. separation of the angle-dependent part
using spherical harmonics. In the process the equations split into
the equations for polar perturbations (even ones) and for axial
perturbations (odd ones). Accelerated reference systems have in
metric coefficient $g_{00}$ a term like $2a r 
cos\theta$, where $a$ is acceleration.
This fact is consistent with the equivalence principle, i.e. in the
accelerated charge rest reference system we have a uniform
gravitational field. So a charge motion dynamics is defined by the
first spherical harmonic for polar perturbations and we will look
for the solutions of the equations for radial functions
corresponding to the first spherical harmonic of polar
perturbations. The initial unperturbed metrics is described by the
Reisner-Nordstrom solution:
\begin{eqnarray}
\label{Nord}
ds^2&=&\Delta/r^2 dt^2-r^2/\Delta dr^2-r^2(\sin^2\theta d\phi^2+d\theta^2), \\
\Delta&=&r^2-2mr+Q^2,\nonumber
\end{eqnarray}

where we assume $c=\kappa=1$. Using these units electron mass and
charge are respectively:

$$m_e=6.67\quad 10^{-56} \mbox{ sm}, e=1.38\quad 10^{-34} \mbox{ sm}$$

Reisner-Nordstrom solution describes a charged black hole when $Q<m$.
When $Q>m$ horizons disappear and we have a naked singularity. We
will be interested in exactly this case since in a classical black
hole metrics a classical radius is under the horizon but the most
interesting things take place exactly at the distances of the order
of the classical radius.
Metrics perturbation corresponding to the first spherical harmonic
of polar perturbations is represented in the following expression:

\begin{eqnarray}
\label{Pert}
h_{\mu\nu}&=&\left(
\begin{array}{cccc}
h_{00}\cos\theta&h_{01}\cos\theta&0&-h_{03}\sin\theta\\
h_{01}\cos\theta&h_{11}\cos\theta&0&-h_{13}\sin\theta\\
0&0&h_{22}\cos\theta \sin^2\theta&0\\
-h_{03}\sin\theta&-h_{13}\sin\theta&0&h_{22}\cos\theta
\end{array}\right)\\
A_\mu&=&(A_0\cos\theta,A_1\cos\theta,0,-A_3\sin\theta)\nonumber
\end{eqnarray}

Linear perturbation theory equations are reduced to a single wave
equation for a gage-invariant (independent of infinitely
small transformations $x^{\mu'}=x^{\mu}+\xi^{\mu}$, where $\xi^{\mu}= (\xi^0(r,t)\cos\theta$,$ \xi^1(r,t)\cos\theta$, $0, -
\xi^3(r,t)\sin\theta)$) function $H_3$, that can be used to
express all the metric coefficient and electromagnetic potentials:
\begin{equation}
\label{base}
-H_{3,tt}+H_{3,r^*r^*}-\frac{2\Delta}{r^2(r-\frac{2}{3}r_0)^2}\left(1-
\frac{2Q^2}{r^2}+\frac{16r_0Q^2}{9r^3}-\frac{4r_0^2Q^2}{9r^4}\right)H_3=0
\end{equation}
Where $r^*$ is defined by $dr/dr^*=\Delta/r^2$, and $r_0=Q^2/m$ is the classical radius. At
that, in gage
\begin{equation}
\label{calib}
h_{22}=h_{03}=A_3=0,\qquad h_{11}=\frac{2r}{\Delta}h_{13}
\end{equation}

the metric coefficients and electromagnetic 4-vector potential are
the simplest:
\begin{eqnarray}
\label{res1}
h_{01}&=&-H_{3,t} \nonumber \\
h_{13}&=&H_3 \nonumber \\
h_{00}&=&-\frac{2\Delta^2}{r^4}H_{3,r}+\left(\frac{\Delta^2+3Q^2\Delta}{r^5}-
\frac{6\Delta Q^2}{r^4r_0}+\frac{\Delta}{r^3}-\frac{4\Delta^2}{r^4(r-\frac{2}{3}r_0)}
\right)H_3\nonumber\\
A_0&=&Q \Delta \left[\frac{2}{r^4} H_3+\left(\frac{3}{2r^2r_0}-
\frac{1}{r^3}\right)H_{3,r}\right]\\
A_1&=&\frac{3 Q r(r-\frac{2}{3}r_0)}{2\Delta r_0}H_{3,t}\nonumber
\end{eqnarray}

Equation (\ref{base}) was obtained for the first time in paper~\cite{Bicak}. In this
paper the time-independent solution corresponding to the
asymptotically uniform electric field $E_0$ has been found. The
corresponding acceleration $a=Q/m E_0$ was exactly equal to Lorenz's
force. If we apply this equation to a point source with the
electron - like parameters and limit ourselves with the distances of
order of an electron classical radius, we can set the
gravitational terms (terms of order $\frac{Q^2}{r_0^2} \sim 10^{-40}$) to zero. At that, $\Delta\to r^2,\quad Q^2\to0$ and equations (\ref{base}), (\ref{res1})
will look as follows:
\begin{equation}
\label{base1}
-H_{3,tt}+H_{3,rr}-\frac{2}{(r-\frac{2}{3}r_0)^2}H_3=0
\end{equation}
\begin{eqnarray}
\label{res2}
h_{00}&=&-2H_{3,r}+\left(-\frac{4}{r-\frac{2}{3}r_0}+\frac{2}{r}\right)H_3\nonumber\\
h_{11}&=&\frac{2}{r}H_3\\
A_0&=&Q\left[\left(\frac{3}{2r_0}-\frac{1}{r}\right)H_{3,r}+\frac{2}{r^2}H_3\right]
\nonumber\\
A_1&=&Q\left(\frac{3}{2r_0}-\frac{1}{r}\right)H_{3,t}\nonumber
\end{eqnarray}
Equations (\ref{base1}) are well known in electrodynamics. The equations for
the electromagnetic potential radial functions look just this way
after variables separation. The difference is only in the fact that
coefficient pole is not in zero but at the two thirds of a
classical radius. It may seem, though, that when the gravitational
terms approaches zero we should get purely electrodynamic
equations (though for an accelerated reference system). The
gravitation disappeared and left an unexpected trail after itself
just as the Cheshire Cat left its smile after itself~\cite{Kar}.
We can write down the general solution of equation~(\ref{base1}):
\begin{equation}
\label{pub}
H_3=(C_1 t+C_2)\left(r-\frac{2}{3}r_0\right)^2+\frac{C_3t+C_4}{r-\frac{2}{3}r_0}+\qquad
\end{equation}
\begin{equation}
\label{pubw}
\qquad +f_1'(r-t)+ \frac{f_1(r-t)}{r-\frac{2}{3}r_0}+
f_2'(r+t)+\frac{f_2(r+t)}{r-\frac{2}{3}r_0}
\end{equation}

If in (\ref{pub}) we set $C_2=\frac{r_0}{3Q}E_0=\frac{Q}{3m}E_0$, and set all other constants to zero, we
will get asymptotically uniform electric field $E_0$ and the
acceleration will be $a=\frac{Q}{m}E_0$. Bicak obtained the same results
by finding the exact solution for equation (\ref{base}). Notice, that there
are no runaway solutions in~(\ref{pub}),(\ref{pubw}) that are so common
for electrodynamics.
Let's consider the wave solutions (\ref{pubw}). They show unlimited growth
when approaching the source and become infinite at the two thirds
of the electron classical radius. This divergence cannot be removed
by taking half a difference of incident and reflected waves as is
usually done in electrodynamics since the pole in field and metrics
expressions will not disappear. It is understandable that at some
perturbation amplitude we have to take into account the
contribution of the perturbation theory higher orders. Anderson in~\cite{Ander} performed decomposition by the charge powers. In our case it
corresponds to the pole decomposition by the powers of $r_0/r$ when
the singularity disappears and we will get radiation reaction force
and runaway solutions related to it.
It's evident that in reality the growing perturbation amplitude is
limited at least by the intensity equal to that of an electron
own field. When this intensity is reached it is necessary to take
into account the contribution of perturbation theory higher orders
that will become equal to the first order contribution. This way GR
makes linear classical electrodynamics irrelevant when dealing with
a charge motion dynamics and field description at the distances of
the order of the classical radius.

\section*{Reasons for Singularity in Solution}

It is useless to take into account the second and higher orders of
perturbation theory if we do not understand the reason giving rise
to the singularity in (\ref{pubw}). By its behavior this singularity is
similar to electromagnetic wave interaction with electrons in
plasma when the wave approaches the critical density where it comes
to resonance with Lengmuire oscillations. But in this case there is
no resonant frequency and it leads to conjecture about the interaction
with some unstable mode, i.e. about the charged source metrics instability.
The issue of black hole stability is a classical issue of
perturbation theory. This issue is dealt with in papers by Regge - Wheeler~\cite{Wheel} (singularity stability in Swartzsield metrics), by
Carter~\cite{Cart} (theorem on Kerr's metrics stability) and by Moncrief~\cite{Moncr}, who has considered the stability of Reisner-Nordstrom black
hole. Moncrief has considered all spherical harmonics of the
rotational and polar perturbations of Reisner-Nordstrom metrics for
the black hole ($Q<m$) outside ($r>r_+$, where $r_+$ is the event horizon)
and shown that there are no exponentially growing solutions. Still,
there is one rotational mode, that he has not considered, and that
can be unstable. Let the perturbation be like that:
\begin{eqnarray}
\label{perta}
h_{\mu\nu}&=&\left(
\begin{array}{cccc}
0&0&h_{02}&0\\
0&0&h_{12}&0\\
0&0&0&0\\
h_{02}&h_{12}&0&0
\end{array}\right)\\
A_\mu&=&(0,0,A_2,0)\nonumber
\end{eqnarray}
If we look for the solutions similar to $h_{\mu\nu}=e^{i\omega t}h_{\mu\nu}(r)\quad A_2=
e^{i\omega t} A_2(r)$ then the first order
perturbation theory equations will look like follows:
\begin{eqnarray}
\label{basea}
\left\{
\begin{array}{rcl}
\omega^2 H_1+H_{1,r^*r^*}+\frac{\Delta(6mr-4Q^2)}{r^6}H_1-\frac{8\Delta Q}{r^5}A_2&=&0\\
\omega^2 A_2+A_{2,r^*r^*}-\frac{4\Delta Q^2}{r^6}A_2+\frac{\Delta Q}{r^5}H_1&=&0
\end{array}
\right.\\
H_1=\frac{h_{02,r}-h_{12,t}}{r^3}-\frac{4 Q A_2}{r^4}\nonumber
\end{eqnarray}
Following the way
Moncrief used to prove Reisner-Nordstrom metrics stability,
one can show, that some of $\omega^2$ eigenvalues are necessery negative.
 That means the presence
of exponentially growing solutions, i.e. instability. The physical
properties of the unstable solution are interesting in relation to
electromagnetic fields. In the unstable solution the magnetic field
is directed along meridians and the electrical one along parallels.
There is no angular dependency and Pointing's vector is everywhere
directed along the radius looking to the sphere with almost
classical radius. So the energy is flowing into the potential well
with the bottom located at the classical radius. To that end it is interesting
to note that in metrics~(\ref{Nord}) the rest point of uncharged test particles is
located at the classical radius. The attracting potential at the
classical radius becomes repulsive, i.e. there really exists a
gravitational potential well in metrics and this well seems to be the
reason for this singularity.
One can agry that this mode~(\ref{perta}) has one significant drawback. It is singular along $Z$
axis. That is, just like in case with Dirac's monopole there exists
a thread along which the magnetic field is infinite. But if we take
into account the higher orders of the perturbation theory this
singularity will disappear. The open question is whether this mode will
still be unstable. One more question is how will this mode
transform if we turn on rotation and move from Reisner-Nordstrom
solution to that by Kerr-Newman. The variable separation for Kerr-
Newman solution is still unknown as well as its
stability. At the same time even a solution with infinitely small
rotation has a different topology. The singularity will change from
point to ring-like. So it can be expected that the singular mode~(\ref{basea}) will become regular. B. Carter (author of the famous Carter's
theorem on Kerr's metrics stability) in his recent paper "Has the
black hole equilibrium problem been solved?"~ \cite{Cart1} says that his
theorem is related only to vacuum-type solutions and the issue of
{\em electrovacuum} solutions is still open.

\section*{Conclusion}

In conclusion it is shown that GR (in contrast with classical electrodynamics)
in linear perturbation theory forecasts the unlimited growth of the
dipole perturbation. This fact is a circumstantial evidence that
the spherically symmetric solution for a charged source is
unstable. It is natural to suppose that the potential well in wave
equation~(\ref{basea}) is related to the gravitational potential well of the
metrics~(\ref{Nord}). There is an analogous potential well in Kerr-Newman
metrics. Let's ask ourselves: at what momentum the potential well in
Kerr-Newman metrics will be substantially different from that in
the spherically symmetric Reisner-Nordstrom metrics? The analysis
of test particles radial trajectories in Kerr-Newman metrics
moving along the axis shows that the potential well is moving away
from the center and becomes smaller in depth only when $a$ (Kerr geometry parameter related to rotation) is greater than $r_0$,when
the momentum $M=mac>mr_0c=Q^2/c$. It leads to conclusion that if potential
well flattening due to rotation can make the metrics stable it can
happen only when momentum $M>Q^2/c$. The real electron momentum $M=\hbar/2=68.5\quad e^2/c$. Since mass is absent in the expression for momentum, then
according to this hypothesis proton momentum can differ from that
of an electron at most in the nineteenth digit (the next term order of magnitude $\sim \frac{m}{Q}=\frac{\sqrt{\kappa}m_Pc^2}{e}\sim 
10^{-18}$).

\section*{Acknowledgements}

Author would like to express his acknowledgement for useful
discussions to M.V.Gorbatenko, V.V.Kassandroff, S.P.Kelner,
V.I.Kogan, V.D.Shafranov and to O.N.Tazetdinov for this paper
translation. I would like to express gratitude to my Teacher
Vladimir Nikolaevich Likhachev.


\begin{thebibliography}{99}

\bibitem{Ander} James L. Anderson. Phys. Rev. D,  56, No. 8, 4675 (1997).
\bibitem{Bicak} J.Bicak, Proc.R.Soc.Lond. A 302,429(1980)
\bibitem{Aut}
M.B.Golubev,VANT Ser.:"Teoreticheskay i prikladnay phizika", No 1, 59(1998)
\bibitem{Moncr}Vincent Moncrief. Phys. Rev. D, 12,No. 6,1526(1975)
\bibitem{Wheel}Tullio Regge, John A. Wheeler, Phys. Rev. 108, No. 4, 1063(1957)
\bibitem{Cart}Carter B. Phys. Rev. Lett., 26, 331-333, 1972
\bibitem{Kar} L.Carrol "Alice in wonderland"
\bibitem{Cart1}B.Carter,"Has the black hole equilibrium problem been solved?"(1997)[gr-
gc/9712038]
\end{thebibliography}
\end{document}